\documentclass[authoryear,5pt]{elsarticle}

\usepackage{amsmath,amssymb,mathtools,mathrsfs}

\usepackage{lineno,hyperref}

\usepackage{fullpage}

\usepackage{amsmath}
\usepackage{amsfonts}
\usepackage{caption}
\usepackage{subcaption}
\usepackage{soul}

\usepackage{natbib}
\usepackage{algorithm}
\usepackage{algpseudocode}%

\usepackage{booktabs}
\usepackage{epstopdf}

\usepackage{physics}
\usepackage[abs]{overpic}
\usepackage{tikz}
\usepackage{wasysym}
\usepackage{relsize}
\usepackage{pifont}
\usepackage{xcolor}
\usepackage{tabularx}
\usepackage{amsmath}
\usepackage{amssymb}
\usepackage{xcolor}
\usepackage{siunitx}
\usepackage{natbib}
\usepackage{amsmath}

\newcommand{\rout}[1]{}

\setcitestyle{authoryear,open={(},close={)}} 

\usepackage[dvipsnames]{xcolor}

\usepackage{graphicx}
\usepackage{float}
\usepackage[english]{babel}
\usepackage{hyperref}
\usepackage{amssymb}
\usepackage[normalem]{ulem}
\usepackage{amsmath}
\usepackage{amsthm}
\usepackage{amsfonts}
\usepackage{enumerate}
\usepackage{mathtools}

\usepackage{lineno}
\usepackage{bm}

\modulolinenumbers[5]

\journal{Elsevier}

\usepackage[dvipsnames,table]{xcolor} 

\graphicspath{{figure/}}

\usepackage{amssymb}

\errorcontextlines\maxdimen

\makeatletter
\newcommand*{\algrule}[1][\algorithmicindent]{\makebox[#1][l]{\hspace*{.5em}\vrule height .75\baselineskip depth .25\baselineskip}}%

\newcount\ALG@printindent@tempcnta
\def\ALG@printindent{%
    \ifnum \theALG@nested>0
        \ifx\ALG@text\ALG@x@notext
            \addvspace{-3pt}
        \else
            \unskip
            \ALG@printindent@tempcnta=1
            \loop
                \algrule[\csname ALG@ind@\the\ALG@printindent@tempcnta\endcsname]%
                \advance \ALG@printindent@tempcnta 1
            \ifnum \ALG@printindent@tempcnta<\numexpr\theALG@nested+1\relax
            \repeat
        \fi
    \fi
    }%
\usepackage{etoolbox}
\patchcmd{\ALG@doentity}{\noindent\hskip\ALG@tlm}{\ALG@printindent}{}{\errmessage{failed to patch}}
\makeatother
\begin{document}

\begin{frontmatter}

\title{Tunable Pathway Selection in Coupled Multistable Snap-Through Systems }

\author{Weicheng Huang$^{1,\dagger, \star}$, Ke Huang$^{2,\dagger}$, Jiaying Zhang$^{2,\star}$, Mingchao Liu$^{3,\star}$,}

\address{
$^{1}$School of Engineering, Newcastle University, Newcastle upon Tyne, NE1 7RU, UK\\
$^{2}$School of Aeronautic Science and Engineering, Beihang University, Beijing, 100191, China\\
$^{3}$Department of Mechanical Engineering, University of Birmingham, Birmingham, B15 2TT, UK\\ 
$^{\dagger}$ W.H. and K.H. contributed equally to this work. \\
$^{\star}$Corresponding authors: weicheng.huang@newcastle.ac.uk (W.H.);\\
$^{\star}$Corresponding authors: jiaying.zhang@buaa.edu.cn (J.Z.); \\
$^{\star}$Corresponding authors: m.liu.2@bham.ac.uk (M.L.).

}

\begin{abstract}

Multistable mechanical systems can store and release elastic energy through snap-through instabilities, but controlling transition pathways between stable states remains challenging when multiple routes are accessible. %
Here, we introduce a two-mass von Mises truss as a general model for studying pathway selection governed by coupled saddle-node bifurcations. 
The system consists of two coupled snap-through units with geometric imperfections, giving rise to four stable configurations: a fully inverted state, a fully natural state, and two intermediate mixed states. 
We show that the coupling stiffness reorganizes the quasi-static bifurcation structure and selects among three transition pathways under release: sequential snapping through one mixed state, direct cooperative snapping, or sequential snapping through the other mixed state. 
Using pseudo-arclength continuation, we track the relevant saddle-node bifurcations and identify the parameter regimes associated with each quasi-static pathway. 
We then demonstrate that dynamic bifurcation delay provides an additional rate-dependent mechanism for pathway selection. 
Even when the quasi-static bifurcation structure favours a unique sequential pathway, finite-rate loading delays snap-through beyond the corresponding static saddle-node points and can reorder the snapping sequence of the two units. 
A local reduction of the coupled dynamics near each saddle-node yields normal forms with coupling-dependent critical points and coefficients. 
The resulting theory identifies distinct rate-dependent delay laws in the inertia-dominated and overdamped regimes and predicts the critical rate at which the snapping order reverses. 
These results establish a general mechanics framework for tuning transition pathways in multistable systems through elastic coupling and loading-rate control.

\end{abstract}


\end{frontmatter}

\section{Introduction}
\label{sec:introduction}

Elastic instabilities have traditionally been treated as failure or serviceability limits in structural design.
More recently, buckling, snap-through, and related bifurcations have been exploited as functional mechanisms for multistability, reversible shape change, rapid energy release, and programmable mechanical response \cite{Reis2015,Kochmann2017,Lu2026}.
Such mechanisms are particularly prevalent in slender structures, including beams, ribbons, plates, and shells, and underpin a broad range of applications in morphing and deployable structures, soft robotics, and mechanical metamaterials \cite{Bertoldi2017,Li2020,Liu2023,yang2023morphing,huang2026tutorial}.
Representative examples include confinement-controlled biholar sheets with programmable hysteretic responses, chains of bistable elements for shock absorption, and interconnected fluidic segments that amplify actuator motion and force
\cite{Florijn2014,Cohen2014,Overvelde2015}.
Related architectures have been developed for reusable energy absorption, mechanical memory and computing, and robotic locomotion \cite{Meng2022,Yang2024Energy,Yasuda2021,Mei2023,Yang2024Kirigami,Guo2024}.
In such systems, instability is not merely a local loss of stability, but can serve as a mechanism for organising transitions between functional configurations.
For multistable structures, the design challenge therefore extends beyond creating multiple stable states to controlling the pathways by which the system transitions between them.

\begin{figure}[!h]
    \centering
    \includegraphics[width=0.85\linewidth]{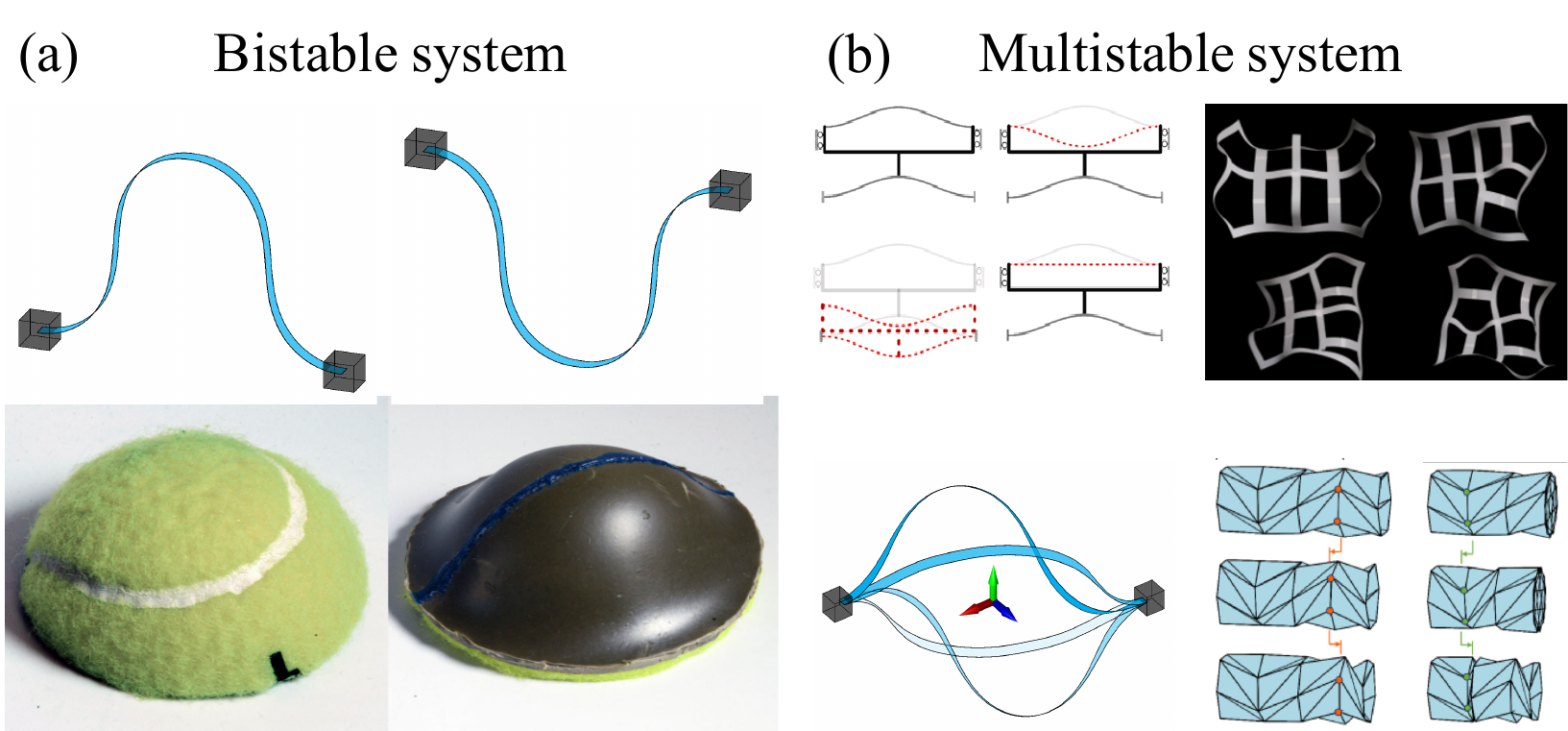}
    \caption{Representative bistable and multistable elastic systems.
    (a) Bistable systems: a compressed buckled ribbon and a shallow-arched jumping popper \cite{Huang2021Ribbon,Pandey2014}.
    (b) Multistable systems: a multistable mechanical metamaterial, a buckled ribbon subjected to transverse shear, a multistable origami structure, and a multistable thin-shell metastructure \cite{Findeisen2017,Yu2019,Zhou2025,huang2024exploiting}.}
    \label{fig:bistable_plot}
\end{figure}

The distinction between bistable and multistable systems is illustrated in Fig.~\ref{fig:bistable_plot}. Figure~\ref{fig:bistable_plot}(a) shows representative bistable systems with two stable configurations, whereas Fig.~\ref{fig:bistable_plot}(b) shows representative multistable systems with several coexisting stable configurations and multiple possible transition pathways.
In an energy-landscape description, stable configurations correspond to local minima separated by energy barriers and saddle points.
As boundary loading reshapes this landscape, a stable minimum can coalesce with a neighbouring saddle in a saddle-node (SN) bifurcation and disappear, triggering snap-through towards another accessible stable state.
For multistable systems whose transitions are governed by competing SN bifurcations, pathway selection therefore depends not only on the available stable states, but also on the ordering of their stability losses and the accessibility of the remaining basins.

Multistability can arise intrinsically from the geometry and mechanics of a single structural architecture or emerge from the interaction of coupled bistable and snap-through building blocks.
The first class includes morphing structures, soft actuators, folded shells, origami systems, ribbon clusters, and prestressed curved rods, in which bending, stretching, hinge kinematics, or stored curvature generate multiple stable configurations \cite{Li2020,Liu2023,Luo2024,Lechenault2015,Xi2023,Meeussen2025,Hong2025RibbonCluster,Leanza2026RodOri}.
The second class includes systems assembled from interacting buckling beams, shells, domes, curved ligaments, and negative-stiffness elements \cite{Findeisen2017,Rafsanjani2015,Wu2023,Huang2025AIAA}.
In such modular systems, the global response is not simply a superposition of independent bistable units.
Elastic interactions modify the local stability of individual units, shift or merge bifurcations, and create collective transition pathways between global states \cite{Yang2023,Huang2026Energy}.
Coupled snap-through units therefore provide a natural setting in which local instabilities compete to organise the global sequence of transitions.

The mechanics of multistable transitions has been studied using models with different levels of abstraction.
Hysteron and transition-graph descriptions represent bistable elements as binary units with state-dependent switching thresholds and are effective for analysing collective switching, memory, and admissible transition sequences \cite{VanHecke2021,Kwakernaak2023,Liu2024Controlled,Shohat2025,Paulsen2026}.
Continuum theories and finite-element models resolve local deformation, stress localisation, and long-range elastic interactions in geometrically complex structures \cite{Liu2023,Findeisen2017}.
Reduced-order mechanical models retain explicit force--displacement relations, equilibrium branches, and stability information while remaining amenable to continuation and energy-landscape analysis \cite{Cohen2014,Roller2024,Huang2026Energy}.
Across these approaches, most strategies for controlling transition pathways remain essentially quasi-static: geometry, stiffness, prestress, imperfections, boundary constraints, and loading sequences are varied to reshape the energy landscape and thereby alter stable states, branch connectivity, critical loads, and bifurcation structure \cite{Li2020,Chen2021Nature,Wu2023,Radisson2023,WuPasini2026}.
For interacting snap-through units, these changes determine which instability occurs first and which intermediate states remain accessible thereafter.
Elastic coupling therefore provides a control parameter for reorganising competing SN bifurcations and selecting among alternative transition pathways.

Finite-rate loading introduces a distinct mechanism for pathway selection.
Whereas quasi-static transitions are organised by equilibrium branches and their loss of stability, dynamic transitions depend on the competition between the imposed loading timescale and the intrinsic timescales of the structure.
Inertia, damping, and internal relaxation can shift the observed instability threshold and alter the selected mode or transition pathway \cite{Gladden2005,Box2020,Kodio2020}.
Near an SN bifurcation, critical slowing down allows the system to remain close to the remnant of the disappearing equilibrium before escaping, producing dynamic bifurcation delay \cite{Gomez2017,Liu2021Delayed}.
When several local instabilities compete, their delays need not be equal, so the corresponding dynamic thresholds may cross even when their quasi-static ordering remains unchanged.
Rate effects can consequently reverse the snapping order, bypass intermediate states, or redirect the transition pathway, as observed in mode skipping, symmetry-dependent transitions, dynamic avalanches, and race conditions \cite{huang2024exploiting,Wang2024Transient,Lindeman2023,Jin2025DynamicAvalanches}.
Loading rate therefore provides a dynamic control parameter that can reorder transitions without altering the underlying quasi-static bifurcation structure.

Against this background, a mechanics framework connecting coupling-controlled pathway selection with rate-dependent reordering of competing local SN instabilities remains lacking.
Hysteron-based models describe transition networks but do not directly resolve the continuous equilibrium branches and SN bifurcations underlying elastic snap-through; bifurcation-based studies resolve these structures but are predominantly quasi-static; and dynamic snap-through analyses commonly focus on a single element or on mode competition within a continuous elastic structure \cite{Huang2026Energy,VanHecke2021,Liu2024Controlled,Radisson2023, Liu2021Delayed,huang2024exploiting}.
To clarify how elastic coupling and loading rate jointly govern pathway selection, we consider a two-mass von Mises truss as a minimal model of two coupled snap-through instabilities.
The model contains two elastically coupled snap-through units and four stable configurations, allowing both sequential and cooperative transitions between the fully inverted and fully natural states.
Pseudo-arclength continuation and energy-landscape analysis are used to resolve the quasi-static equilibrium branches, their stability, and the accessibility of the intermediate mixed states.
Full two-degree-of-freedom dynamic simulations are then combined with continuation of the coupled SN points and local reductions of the dynamics near each SN bifurcation.
The results show that coupling stiffness reorganises the competing SN thresholds and selects the quasi-static transition pathway, whereas finite-rate release generates unequal bifurcation delays that can reorder the snapping sequence.
The resulting coupling-dependent local theory predicts the dynamic snap-through thresholds and the critical rate for pathway switching, thereby connecting the quasi-static bifurcation structure to rate-controlled pathway selection.

The remainder of this paper is organised as follows.
Section~\ref{sec:problem_setup} defines the stable states and possible transition pathways.
Section~\ref{sec:trussmodel} develops the two-mass von Mises truss model and its non-dimensional governing equations.
Section~\ref{sec:static_tunable_path} examines the equilibrium branches, stability, and coupling-controlled quasi-static pathway selection.
Section~\ref{sec:dynamic_result} analyses finite-rate release, develops the local SN reductions, and constructs the rate-controlled pathway map.
Section~\ref{sec:Discussion} discusses the connections between the general truss model and real multistable mechanical systems.
Finally, Section~\ref{sec:conclusion} summarises the main findings and the future implications for programmable multistable systems.

\section{Problem setup}
\label{sec:problem_setup}

\begin{figure}[!h]
    \centering
    \includegraphics[width=\linewidth]{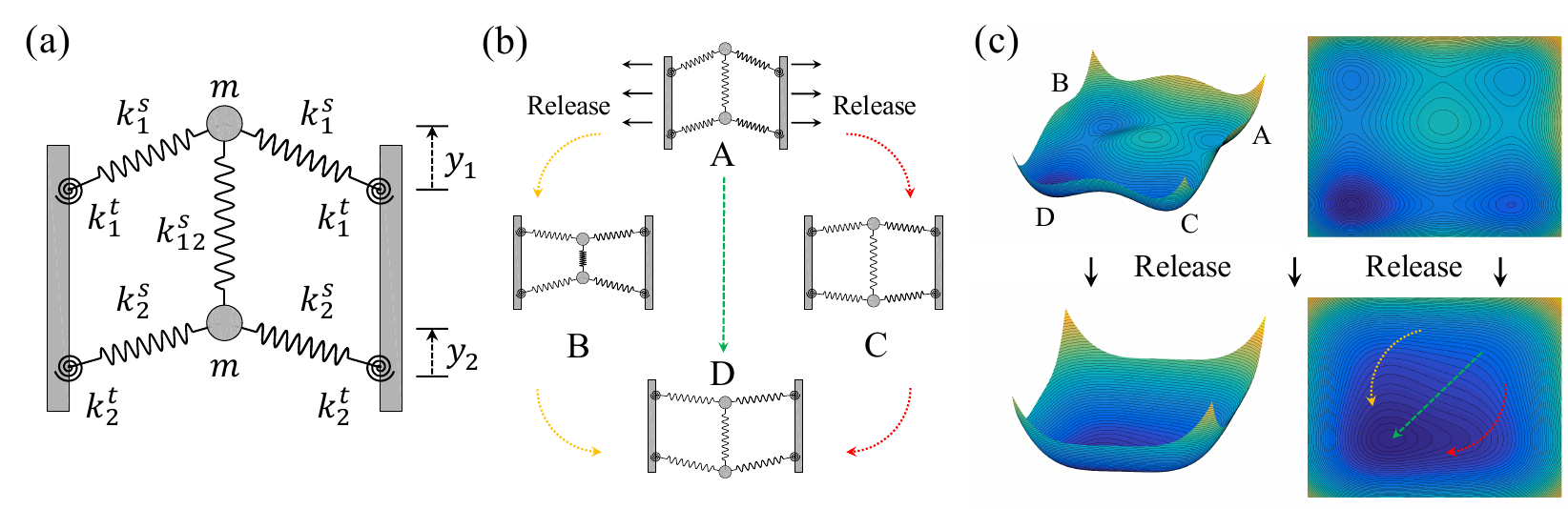}
    \caption{Two-mass von Mises truss and transition pathways.
    (a) Two-degree-of-freedom truss consisting of two vertically constrained masses, two bistable snap-through units, and an elastic coupling spring of stiffness $k_{12}^{s}$.
    (b) Four stable configurations: the fully inverted state $A$, the fully natural state $D$, and the two mixed states $B$ and $C$, together with the three possible transition pathways from $A$ to $D$ under release.
    (c) Energy-landscape representation of pathway selection during release, illustrating direct and sequential transitions between stable states.}
    \label{fig:model_plot}
\end{figure}

We consider the two-degree-of-freedom two-mass von Mises truss shown in Fig.~\ref{fig:model_plot}(a) as a general system for studying transition-pathway selection in coupled snap-through structures.
The system consists of two point masses constrained to move vertically, with displacements $y_{1}$ and $y_{2}$, and is driven by the prescribed boundary separation $\delta$.
Geometric asymmetry is introduced through the stress-free angles $\alpha_{1}$ and $\alpha_{2}$ of the rotational springs, breaking the up--down symmetry and distinguishing the inverted and natural configurations.
As $\delta$ is increased, the structure is released from an inverted configuration towards a natural configuration and may undergo one or more snap-through transitions along the way.
This release protocol provides a specific setting in which to determine which transition pathway is selected, how elastic coupling controls the quasi-static pathway, and how finite loading rate can reorder the transitions dynamically.

Throughout this work, we focus on negative geometric imperfections, $\alpha_{1}<0$ and $\alpha_{2}<0$.
Under this convention, the fully inverted configuration corresponds to positive displacements of both degrees of freedom, whereas the fully natural configuration corresponds to negative displacements of both degrees of freedom.
The four stable configurations are classified according to the signs of $(y_{1},y_{2})$ as\begin{equation}
A:(y_{1}>0,y_{2}>0),\quad
B:(y_{1}<0,y_{2}>0),\quad
C:(y_{1}>0,y_{2}<0),\quad
D:(y_{1}<0,y_{2}<0).
\label{eq:four_state_classification}
\end{equation}
Here, $A$ denotes the fully inverted state and $D$ the fully natural state, while $B$ and $C$ are mixed states in which only one degree of freedom has snapped through.
In the following, a \textit{transition} refers to an individual snap-through event between stable states, whereas a \textit{transition pathway} refers to the sequence of transitions connecting the initial and final states, together with the associated equilibrium branches and dynamic trajectory.

This classification gives rise to three possible transition pathways from the inverted state $A$ to the natural state $D$:
\begin{equation}
A\rightarrow B\rightarrow D,\qquad
A\rightarrow D,\qquad
A\rightarrow C\rightarrow D .
\label{eq:possible_static_paths}
\end{equation}
The first and third pathways consist of two sequential snap-through transitions with opposite snapping orders, whereas the second corresponds to a direct cooperative transition in which the system reaches $D$ without being arrested in either mixed state, as shown in Figs.~\ref{fig:model_plot}(b),

The problem addressed in this paper is therefore twofold: first, to determine how elastic coupling selects among these pathways under quasi-static release; and second, to establish how finite-rate release modifies the dynamic instability thresholds and can thereby delay, bypass, or reorder the corresponding transitions.
The four stable configurations and their energy-landscape interpretation are illustrated in Figs.~\ref{fig:model_plot}(b) and \ref{fig:model_plot}(c), respectively.

\section{Two-mass von Mises truss model}
\label{sec:trussmodel}

\subsection{Model development}

We employ a two-mass von Mises truss as a general mechanical model for studying how elastic coupling reorganises competing SN bifurcations and how finite-rate loading modifies transition-pathway selection.

As illustrated in Fig.~\ref{fig:model_plot}(a), the upper mass is connected to the lateral boundaries by two identical linear stretching springs, each with natural length $l_0$ and stiffness $k_{1}^{s}$.
Two boundary torsional springs, each with stiffness $k_{1}^{t}$, resist rotation of the corresponding inclined members.
The lower mass is connected to the boundaries in the same manner: its inclined stretching springs have natural length $l_0$ and stiffness $k_{2}^{s}$, while the associated boundary torsional springs have stiffness $k_{2}^{t}$.
The two masses are separated by a reference vertical distance $l_0$ and coupled by an additional linear stretching spring with natural length $l_0$ and stiffness $k_{12}^{s}$.
Both point masses have mass $m$ and are subjected to identical linear viscous damping with coefficient $c$. Their vertical displacements relative to the reference configuration are denoted by $y_1$ and $y_2$.
The prescribed parameter $\delta$ denotes the horizontal distance from each mass to the corresponding lateral boundary. 
The system is initially held at a small value of $\delta$ and subsequently released by gradually increasing $\delta$.
The current lengths of the inclined springs associated with the upper and lower masses, together with that of the coupling spring, are\begin{equation}
l_1=\sqrt{\delta^2+y_1^2},
\qquad
l_2=\sqrt{\delta^2+y_2^2},
\qquad
l_3=l_0+y_1-y_2 .
\label{eq:dimensional_lengths}
\end{equation}
Under the adopted signed-angle convention, the rotations of the inclined members are\begin{equation}
\theta_1=\arctan\left(\frac{y_1}{\delta}\right),
\qquad
\theta_2=\arctan\left(\frac{y_2}{\delta}\right).
\label{eq:dimensional_angles}
\end{equation}
The boundary torsional springs are stress free when the corresponding inclined members attain the prescribed angles $\alpha_1$ and $\alpha_2$, respectively.
Because each truss unit contains two identical inclined stretching springs and two identical boundary torsional springs, the total dimensional potential energy is
\begin{equation}
U(y_1,y_2;\delta)=
k_{1}^{s}\left(l_1-l_0\right)^2
+
k_{2}^{s}\left(l_2-l_0\right)^2
+
\frac{1}{2}k_{12}^{s}\left(l_3-l_0\right)^2
+
k_{1}^{t}\left(\theta_1-\alpha_1\right)^2
+
k_{2}^{t}\left(\theta_2-\alpha_2\right)^2 .
\label{eq:dimensional_energy}
\end{equation}
The equations of motion are therefore
\begin{align}
m\frac{\mathrm{d}^2 y_1}{\mathrm{d}t^2}
+
c\frac{\mathrm{d}y_1}{\mathrm{d}t}
+
\frac{\partial U}{\partial y_1}
&=0,
\label{eq:dimensional_eom_y1}
\\
m\frac{\mathrm{d}^2 y_2}{\mathrm{d}t^2}
+
c\frac{\mathrm{d}y_2}{\mathrm{d}t}
+
\frac{\partial U}{\partial y_2}
&=0,
\label{eq:dimensional_eom_y2}
\end{align}
where $m$ is the mass of the point and $c$ is the associated damping coefficient.

\subsection{Nondimensionalization}

We next introduce the dimensionless variables
\begin{equation}
Y_1=\frac{y_1}{l_0},
\qquad
Y_2=\frac{y_2}{l_0},
\qquad
\Delta=\frac{\delta}{l_0},
\qquad
L_{1}  = \frac {l_{1}}{l_{0}},
\qquad
L_{2}  = \frac {l_{2}}{l_{0}},
\qquad
L_{3}  = \frac {l_{3}}{l_{0}},
\label{eq:dimensionless_variables}
\end{equation}
together with the dimensionless stiffness parameters
\begin{equation}
K_{2}^{s}=\frac{k_{2}^{s}}{k_{1}^{s}},
\qquad
K_{12}^{s}=\frac{k_{12}^{s}}{k_{1}^{s}},
\qquad
K_{1}^{t}=\frac{k_{1}^{t}}{k_{1}^{s}l_0^2},
\qquad
K_{2}^{t}=\frac{k_{2}^{t}}{k_{1}^{s}l_0^2},
\label{eq:dimensionless_parameters}
\end{equation}
By definition, the dimensionless axial stiffness of unit 1 is $K_1^s=1$.
The dimensionless time and damping parameter are
\begin{equation}
T=t\sqrt{\frac{k_{1}^{s}}{m}},
\qquad
\Upsilon=\frac{c}{\sqrt{k_{1}^{s}m}}.
\label{eq:dimensionless_lengths}
\end{equation}
Scaling the total potential energy by $k_1^s l_0^2$, we define the dimensionless potential energy as
\begin{equation}
\mathcal{U}(Y_1,Y_2;\Delta)
=
\left(L_1-1\right)^2
+
K_{2}^{s}\left(L_2-1\right)^2
+
\frac{1}{2}K_{12}^{s}(Y_1-Y_2)^2
+
K_{1}^{t}\left(\theta_1-\alpha_1\right)^2
+
K_{2}^{t}\left(\theta_2-\alpha_2\right)^2 .
\label{eq:normalized_energy_compact}
\end{equation}
The exact dimensionless equations of motion associated with $\mathcal{U}$ are
\begin{align}
\frac{\mathrm{d}^2Y_1}{\mathrm{d}T^2}
+
\Upsilon\frac{\mathrm{d}Y_1}{\mathrm{d}T}
+
\frac{\partial \mathcal{U}}{\partial Y_1}
&=0,
\label{eq:norm_eom_y1}
\\
\frac{\mathrm{d}^2Y_2}{\mathrm{d}T^2}
+
\Upsilon\frac{\mathrm{d}Y_2}{\mathrm{d}T}
+
\frac{\partial \mathcal{U}}{\partial Y_2}
&=0.
\label{eq:norm_eom_y2}
\end{align}
For the release protocol considered here, the dimensionless horizontal distance is prescribed to increase linearly with time as
\begin{equation}
\Delta(T)=\Delta_0+\dot{\Delta}T,
\qquad
\dot{\Delta}>0,
\label{eq:loading_protocol}
\end{equation}
where $\Delta_0$ denotes the initial dimensionless horizontal distance and $\dot{\Delta}$ is the imposed dimensionless release rate.

\subsection{Fourth-order expansion of the potential energy}

For small transverse displacements, $|Y_1|\ll 1$ and $|Y_2|\ll 1$, we expand the dimensionless potential energy to fourth order in $Y_1$ and $Y_2$:
\begin{align}
\mathcal{U}_4(Y_1,Y_2;\Delta)
&=
C_{00}(\Delta)
+
C_{10}(\Delta)Y_1
+
C_{01}(\Delta)Y_2
\nonumber\\
&\quad
+
C_{20}(\Delta)Y_1^2
+
C_{11}(\Delta)Y_1Y_2
+
C_{02}(\Delta)Y_2^2
\nonumber\\
&\quad
+
C_{30}(\Delta)Y_1^3
+
C_{03}(\Delta)Y_2^3
+
C_{40}(\Delta)Y_1^4
+
C_{04}(\Delta)Y_2^4,
\label{eq:reduced_energy_2d}
\end{align}
where the coefficients are
\begin{align}
C_{00}(\Delta)
&=
\left(1+K_{2}^{s}\right)(\Delta-1)^2
+
K_{1}^{t}\alpha_1^2
+
K_{2}^{t}\alpha_2^2,
\label{eq:C00}
\\
C_{10}(\Delta)
&=
-\frac{2K_{1}^{t}\alpha_1}{\Delta},
\label{eq:C10}
\\
C_{01}(\Delta)
&=
-\frac{2K_{2}^{t}\alpha_2}{\Delta},
\label{eq:C01}
\\
C_{20}(\Delta)
&=
\frac{\Delta-1}{\Delta}
+
\frac{1}{2}K_{12}^{s}
+
\frac{K_{1}^{t}}{\Delta^2},
\label{eq:C20}
\\
C_{11}(\Delta)
&=
-K_{12}^{s},
\label{eq:C11}
\\
C_{02}(\Delta)
&=
\frac{K_{2}^{s}(\Delta-1)}{\Delta}
+
\frac{1}{2}K_{12}^{s}
+
\frac{K_{2}^{t}}{\Delta^2},
\label{eq:C02}
\\
C_{30}(\Delta)
&=
\frac{2K_{1}^{t}\alpha_1}{3\Delta^3},
\label{eq:C30}
\\
C_{03}(\Delta)
&=
\frac{2K_{2}^{t}\alpha_2}{3\Delta^3},
\label{eq:C03}
\\
C_{40}(\Delta)
&=
\frac{1}{4\Delta^3}
-
\frac{2K_{1}^{t}}{3\Delta^4},
\label{eq:C40}
\\
C_{04}(\Delta)
&=
\frac{K_{2}^{s}}{4\Delta^3}
-
\frac{2K_{2}^{t}}{3\Delta^4}.
\label{eq:C04}
\end{align}
The fourth-order potential $\mathcal{U}_4$ is used for the quasi-static bifurcation analysis and the local normal-form derivation, whereas the full dynamic simulations employ the exact potential $\mathcal{U}$.
For the quasi-static analysis of the reduced model, the equilibrium configurations are determined by
\begin{equation}
\frac{\partial \mathcal{U}_4}{\partial Y_1}=0,
\qquad
\frac{\partial \mathcal{U}_4}{\partial Y_2}=0.
\label{eq:static_equilibrium_conditions}
\end{equation}
The local stability of each equilibrium is determined from the Hessian matrix $\boldsymbol{\mathcal{K}}$ of the reduced potential energy,
\begin{equation}
\boldsymbol{\mathcal{K}}
=
\begin{bmatrix}
\displaystyle
\frac{\partial^2 \mathcal{U}_4}{\partial Y_1^2}
&
\displaystyle
\frac{\partial^2 \mathcal{U}_4}{\partial Y_1\partial Y_2}
\\[2mm]
\displaystyle
\frac{\partial^2 \mathcal{U}_4}{\partial Y_2\partial Y_1}
&
\displaystyle
\frac{\partial^2 \mathcal{U}_4}{\partial Y_2^2}
\end{bmatrix}.
\label{eq:stability_hessian}
\end{equation}
An equilibrium corresponds to a locally stable configuration when $\boldsymbol{\mathcal{K}}$ is positive definite.
The loss of stability associated with an SN bifurcation is identified when Eq.~\eqref{eq:static_equilibrium_conditions}is satisfied and one eigenvalue of $\boldsymbol{\mathcal{K}}$ vanishes while the other remains positive.

\section{Quasi-static pathway selection by elastic coupling}
\label{sec:static_tunable_path}

We first examine the quasi-static bifurcation structure of the two-degree-of-freedom truss model.
The equilibrium configurations are obtained by solving the stationary conditions in Eq.~\eqref{eq:static_equilibrium_conditions} using pseudo-arclength continuation.
This allows both stable and unstable equilibrium branches to be traced through limit points, revealing the bifurcation structure underlying the snap-through transitions.
Unless otherwise stated, the dimensionless parameters are chosen as
\begin{equation}
K_{2}^{s}=20.0,\qquad
K_{1}^{t}=0.1,\qquad
K_{2}^{t}=2.0,\qquad
\alpha_{1}=-0.04,\qquad
\alpha_{2}=-0.01.
\label{eq:static_parameters}
\end{equation}
Here, $K_1^{s}\equiv1$ by nondimensionalisation.

Since nondimensionalisation does not change the signs of the vertical displacements, we retain the state classification introduced in Section~\ref{sec:problem_setup} (Eq.~\ref{eq:four_state_classification}): $A$ and $D$ denote the fully inverted and fully natural states, respectively, while $B$ and $C$ denote the two mixed states.

Starting from the compressed inverted state $A$ at $\Delta_{0}=0.5$, we increase the dimensionless horizontal distance $\Delta$ quasi-statically.
As $\Delta$ increases, the system successively loses stability and transitions towards the natural state $D$ (see Fig.~\ref{fig:model_plot}b).
Depending on the coupling stiffness, three distinct transition pathways arise:\begin{equation}
\mathrm{(i)}: \; A\rightarrow B\rightarrow D,\qquad
\mathrm{(ii)}: \; A\rightarrow D,\qquad
\mathrm{(iii)}: \;  A\rightarrow C\rightarrow D .
\label{eq:static_paths}
\end{equation}
The first and third pathways correspond to sequential snapping of the two truss units in opposite orders, whereas the second corresponds to a direct cooperative transition from $A$ to $D$ without arrest in either mixed state.
The coupling stiffness $K_{12}^{s}$ therefore acts as a structural control parameter that reorganises the competing bifurcations and selects the quasi-static transition pathway.

\begin{figure}[!h]
    \centering
    \includegraphics[width=0.9\linewidth]{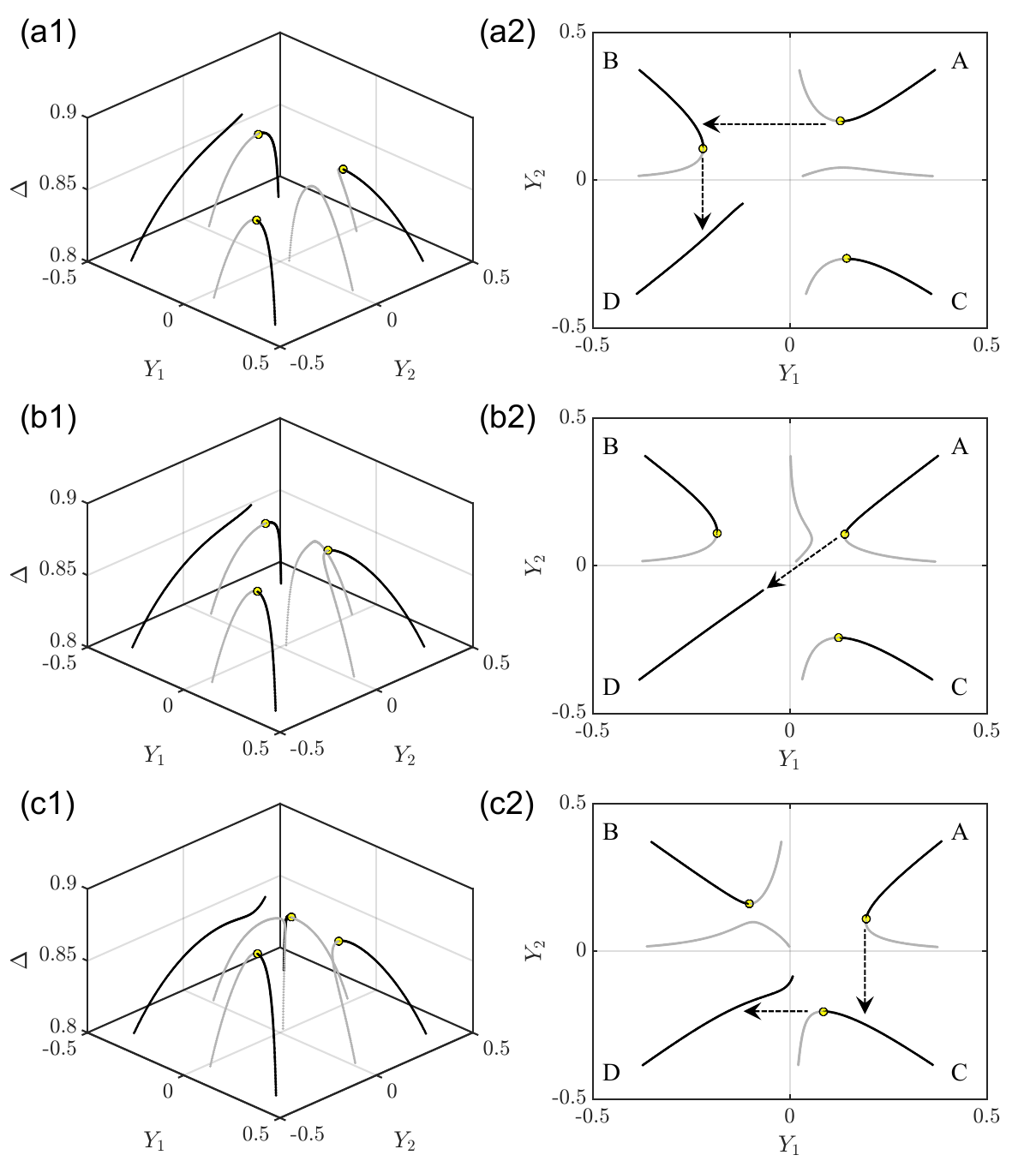}
    \caption{Bifurcation diagrams and transition pathways for different coupling stiffnesses.
    (a1,b1,c1) Three-dimensional equilibrium branches in the $(Y_1,Y_2,\Delta)$ space for $K_{12}^{s}=0.004$, $0.007$, and $0.010$, respectively.
    (a2,b2,c2) Corresponding projections onto the $(Y_1,Y_2)$ plane, with arrows indicating the transitions between stable states during quasi-static release.
    Black and light-grey curves denote stable and unstable equilibrium branches, respectively, and circles mark the saddle-node (SN) bifurcation points.}
    \label{fig:static_plot}
\end{figure}

When the coupling spring is sufficiently weak, the two truss units interact only weakly and behave almost independently.
For the parameters in Eq.~\eqref{eq:static_parameters}, the uncoupled instability thresholds satisfy $(\Delta_c)_1<(\Delta_c)_2$, owing to the unequal stretching stiffnesses, torsional stiffnesses, and stress-free angles of the two units.
The system therefore follows the sequential pathway $A\rightarrow B\rightarrow D$.
A representative bifurcation diagram for this regime is shown in Figs.~\ref{fig:static_plot}(a1) and \ref{fig:static_plot}(a2) for $K_{12}^{s}=0.004$, showing the three-dimensional equilibrium branches and their projection onto the $(Y_1,Y_2)$ plane, respectively.
The first SN bifurcation corresponds to the loss of stability of state $A$, after which the system transitions to the mixed state $B$.
Upon further release, state $B$ loses stability at a second SN bifurcation, leading to the final transition to state $D$.

At intermediate coupling stiffness, the two truss units no longer snap independently.
The instability of one unit is transmitted sufficiently strongly through the coupling spring that the system undergoes a direct cooperative transition from the fully inverted state to the fully natural state, $A\rightarrow D$.
Figures~\ref{fig:static_plot}(b1) and \ref{fig:static_plot}(b2) show this behaviour for $K_{12}^{s}=0.007$.
In this regime, the mixed states are bypassed along the selected quasi-static pathway, even though their equilibrium branches remain present in the full bifurcation structure.

As the coupling stiffness is increased further, the relative ordering of the competing instabilities is reversed.
The coupling shifts the stability thresholds associated with the two local snapping modes such that state $A$ first loses stability towards the mixed state $C$ rather than $B$.
The system therefore follows the alternative sequential pathway $A\rightarrow C\rightarrow D$.
This regime is illustrated in Figs.~\ref{fig:static_plot}(c1) and \ref{fig:static_plot}(c2) for $K_{12}^{s}=0.010$.
Thus, tuning a single elastic coupling parameter switches the quasi-static response between two sequential pathways with opposite snapping orders and a direct cooperative transition.

\begin{figure}[!h]
    \centering
    \includegraphics[width=0.7\linewidth]{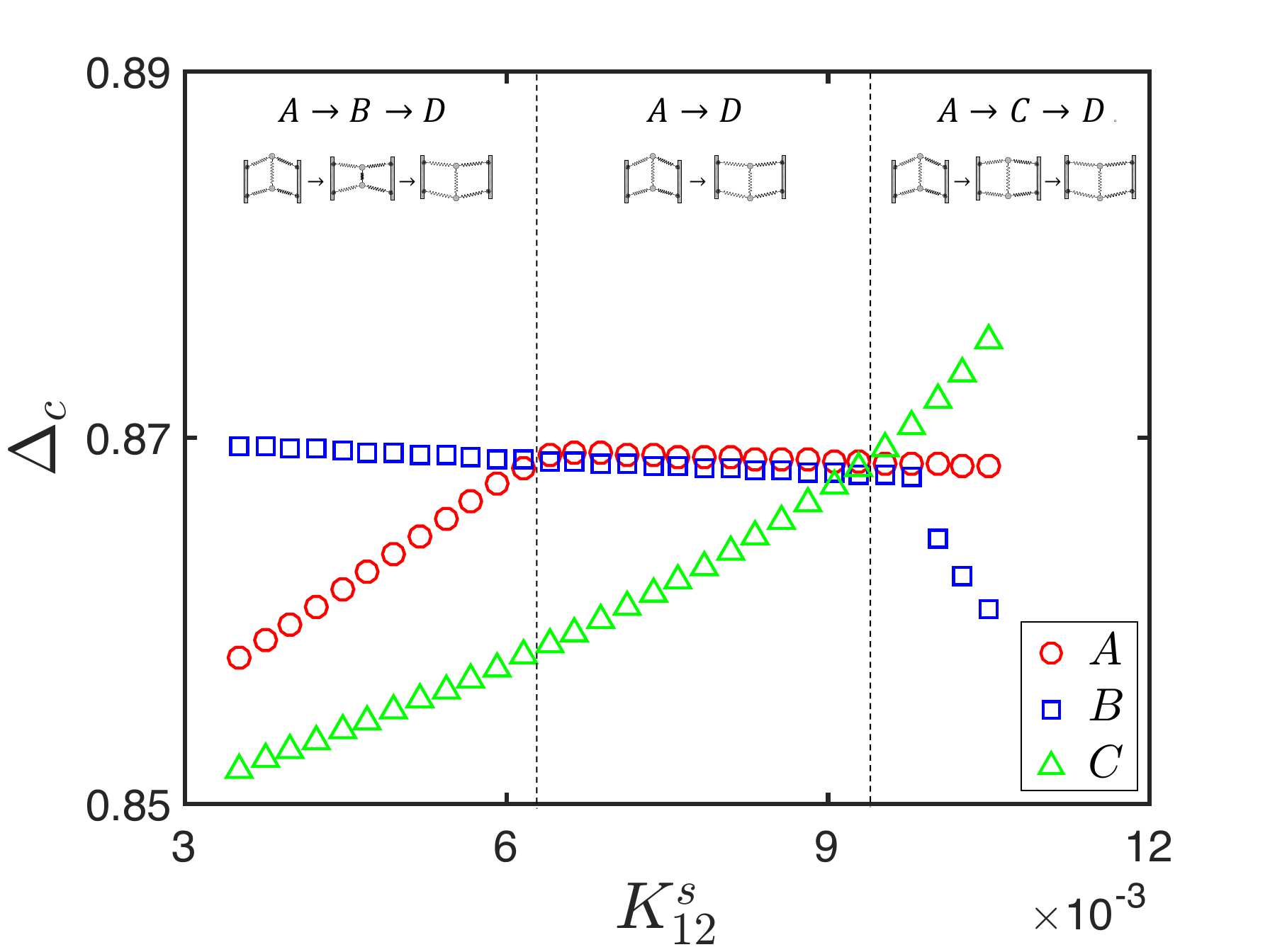}
    \caption{Critical stability thresholds and pathway-selection regimes.
    Critical values $\Delta_c$ at which states $A$, $B$, and $C$ lose stability through saddle-node (SN) bifurcations, plotted as functions of the coupling stiffness $K_{12}^{s}$.
    The relative ordering of these thresholds divides the parameter space into three quasi-static transition regimes, corresponding to the pathways $A\rightarrow B\rightarrow D$, $A\rightarrow D$, and $A\rightarrow C\rightarrow D$ under release.}
    \label{fig:static_plot_2}
\end{figure}

To identify the parameter regimes associated with the three pathways, we perform a parameter sweep over $4\times10^{-3}\leq K_{12}^{s}\leq12\times10^{-3}$.
For each value of $K_{12}^{s}$, we determine the critical value $\Delta_c$ at which states $A$, $B$, and $C$ lose stability through SN bifurcations.
The resulting stability thresholds are plotted in Fig.~\ref{fig:static_plot_2}.
Because the system starts from state $A$ and $\Delta$ increases monotonically, the first transition occurs when the $A$ branch reaches its critical threshold.
The subsequent pathway is then determined by the stability of the mixed states $B$ and $C$ at this same value of $\Delta$.
If $\Delta_c^{A}<\Delta_c^{B}$ while $\Delta_c^{A}>\Delta_c^{C}$, state $B$ remains stable when $A$ loses stability, whereas state $C$ has already lost stability, leading to the sequential pathway $A\rightarrow B\rightarrow D$.
Conversely, if $\Delta_c^{A}<\Delta_c^{C}$ while $\Delta_c^{A}>\Delta_c^{B}$, state $C$ remains stable and the system follows $A\rightarrow C\rightarrow D$.
If $\Delta_c^{A}$ exceeds the stability thresholds of both mixed states, neither $B$ nor $C$ remains available to arrest the transition, and the system undergoes the direct cooperative transition $A\rightarrow D$.
The crossings of the critical curves therefore mark the boundaries between the three quasi-static pathway-selection regimes.
These results show that elastic coupling does more than shift individual instability thresholds: it changes their relative ordering and thereby selects the transition pathway.

\section{Dynamic response and rate-controlled pathway selection}
\label{sec:dynamic_result}

\subsection{Rate-controlled pathway reordering}

We next examine the dynamic response of the two-mass von Mises truss and show that the release rate $\dot{\Delta}$ provides an additional control parameter for transition-pathway selection.
In the quasi-static limit, the selected pathway is governed by the relative ordering of the static stability thresholds. At finite release rates, however, the system can remain transiently near the ghost of a vanished equilibrium after the corresponding static SN bifurcation has been crossed.
This dynamic bifurcation delay shifts the observed snap-through thresholds and can reorder the snapping sequence of the two units, thereby selecting pathways that differ from the quasi-static response.
In this section, we take
\begin{equation}
K_{12}^{s}=0.002,\qquad \Upsilon=0.2,
\end{equation}
while all other parameters are kept the same as in the quasi-static analysis.
For this coupling stiffness, the quasi-static pathway is $A\rightarrow B\rightarrow D$.
The release parameter is prescribed as a linear ramp,
\begin{equation}
\Delta(T)=\Delta_{0}+\dot{\Delta}T,
\end{equation}
where $\dot{\Delta}$ is the dimensionless release rate.
The dimensionless equations of motion, Eqs.~\eqref{eq:norm_eom_y1} and \eqref{eq:norm_eom_y2}, are integrated using the \texttt{MATLAB ODE45} solver.
The system is initialised on the fully inverted equilibrium branch $A$ at $\Delta=\Delta_{0}$.

\begin{figure}[!h]
    \centering
    \includegraphics[width=\linewidth]{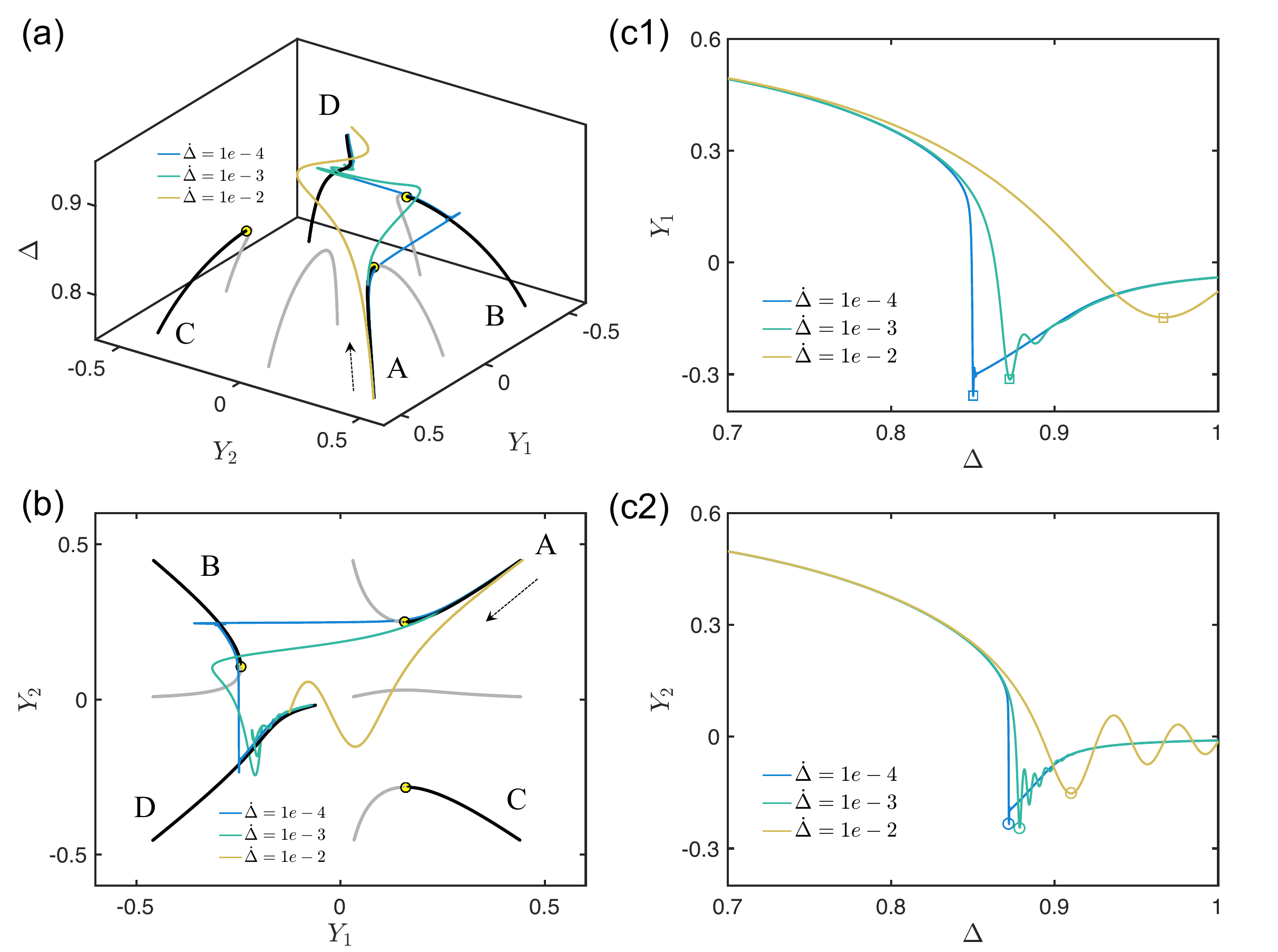}
    \caption{Rate-controlled reordering of transition pathways in the coupled snap-through system.
    (a) Three-dimensional dynamic trajectories for different release rates, $\dot{\Delta}\in\{10^{-4},10^{-3},10^{-2}\}$.
    (b) Corresponding projections onto the $(Y_1,Y_2)$ plane, showing the pathway reordering from $A\rightarrow B\rightarrow D$ to $A\rightarrow D$ and then to $A\rightarrow C\rightarrow D$ as the release rate increases.
    (c1,c2) Evolution of $Y_1$ and $Y_2$ with $\Delta$, respectively.
    Circular and square markers indicate the dynamically identified snapping points of units 1 and 2, respectively.
    The corresponding dynamic processes are shown in Supplementary Movies S1--S3.}
    \label{fig:dynamic_plot}
\end{figure}

For these parameters, the quasi-static bifurcation structure predicts the sequential pathway $A\rightarrow B\rightarrow D$.
Under quasi-static release, unit 1 loses stability first, and the system transitions to the mixed state $B$ before the subsequent snap-through of unit 2.
The dynamic simulations, however, show that the selected pathway changes as the release rate increases.
As shown in Figs.~\ref{fig:dynamic_plot}(a) and \ref{fig:dynamic_plot}(b), increasing $\dot{\Delta}$ progressively alters the trajectory in the $(Y_{1},Y_{2})$ plane and ultimately reverses the snapping order of the two units.

For slow release, for example $\dot{\Delta}=10^{-4}$, the dynamic response remains close to the quasi-static equilibrium pathway.
The system first leaves the fully inverted state $A$, transitions to the mixed state $B$, and subsequently reaches the natural state $D$.
The resulting pathway is therefore consistent with the quasi-static prediction, $A\rightarrow B\rightarrow D$.
As the release rate increases, dynamic bifurcation delay shifts the snap-through events to larger values of $\Delta$.
At an intermediate rate, for example $\dot{\Delta}=10^{-3}$, neither mixed state arrests the motion and the system undergoes a direct cooperative transition, $A\rightarrow D$.
With a further increase in release rate, the relative ordering of the dynamically delayed snapping events is reversed.
For $\dot{\Delta}=10^{-2}$, unit 2 snaps before unit 1, and the system follows the alternative sequential pathway $A\rightarrow C\rightarrow D$.
Thus, although the quasi-static bifurcation structure favours $A\rightarrow B\rightarrow D$, finite-rate release successively reorganises the transition pathway from $A\rightarrow B\rightarrow D$ to $A\rightarrow D$ and ultimately to $A\rightarrow C\rightarrow D$.

Figures~\ref{fig:dynamic_plot}(c1) and \ref{fig:dynamic_plot}(c2) show the corresponding responses of $Y_{1}$ and $Y_{2}$ for $\dot{\Delta}\in{10^{-4},10^{-3},10^{-2}}$ as functions of the release parameter $\Delta$.
To quantify the snapping sequence, we define the dynamic snapping points $\Delta_{\mathrm{snap},1}^{\mathrm{dyn}}$ and $\Delta_{\mathrm{snap},2}^{\mathrm{dyn}}$ as the values of $\Delta$ at which the corresponding dynamic trajectories attain their local extrema during snap-through.
These points are indicated by the circular and square markers in Figs.~\ref{fig:dynamic_plot}(c1) and \ref{fig:dynamic_plot}(c2), respectively, and provide a practical measure of the dynamic snapping order.
For slow release, $\Delta_{\mathrm{snap},1}^{\mathrm{dyn}}<\Delta_{\mathrm{snap},2}^{\mathrm{dyn}}$, indicating that unit 1 snaps first.
For fast release, this ordering is reversed, with $\Delta_{\mathrm{snap},2}^{\mathrm{dyn}}<\Delta_{\mathrm{snap},1}^{\mathrm{dyn}}$.
The release rate therefore acts as a dynamic control parameter that can switch the system between distinct transition pathways without changing its underlying elastic structure.
The corresponding transient dynamics are shown in Supplementary Movies S1--S3.

These results demonstrate that pathway selection in a multistable system is not governed solely by the quasi-static bifurcation structure.
Instead, the observed pathway results from the competition between the intrinsic timescales governing instability growth and the externally imposed loading timescale.
Under slow release, the system closely follows the quasi-static sequence.
As the release rate increases, unequal dynamic bifurcation delays shift the effective snapping thresholds and can eventually reverse their ordering.
This rate-controlled reordering provides a dynamic mechanism for selecting transition pathways in coupled multistable systems.

\subsection{Dynamic bifurcation theory}
\label{sec:coupled_local_normal_form}

\subsubsection{Coupled local reduction}
We next derive a local description of the dynamics near the relevant SN bifurcations of the coupled two-degree-of-freedom system.
Let
\begin{equation}
\mathbf{Y}
=
[
Y_1,Y_2
]^{\mathsf T}
\qquad
\mathbf{G}(\mathbf{Y};\Delta,K_{12}^{s})
=
\frac{\partial \mathcal{U}_{4}}{\partial \mathbf{Y}},
\qquad
\boldsymbol{\mathcal{K}}
=
\frac{\partial \mathbf{G}}{\partial \mathbf{Y}},
\label{eq:coupled_vector_residual}
\end{equation}
where $\mathcal{U}_4$ is the fourth-order potential energy and $\mathcal{K}$ is its Hessian, as defined in Eqs.~\eqref{eq:reduced_energy_2d} and \eqref{eq:stability_hessian}.
The $i$-th coupled SN point, with $i\in\{1,2\}$, satisfies
\begin{equation}
\mathbf{G}
\left(
\mathbf{Y}_{c,i};
\Delta_{c,i},
K_{12}^{s}
\right)
=
\mathbf{0},
\qquad
\boldsymbol{\mathcal{K}}_{c,i}\boldsymbol{\psi}_i
=
\mathbf{0},
\label{eq:coupled_sn_conditions}
\end{equation}
where
\begin{equation}
\boldsymbol{\mathcal{K}}_{c,i}
=
\boldsymbol{\mathcal{K}}
\left(
\mathbf{Y}_{c,i};
\Delta_{c,i},
K_{12}^{s}
\right),
\end{equation}
and $\boldsymbol{\psi}_i$ is the normalized null eigenvector,
$\boldsymbol{\psi}_i^{\mathrm{T}}\boldsymbol{\psi}_i=1$.
Near the SN point, the local displacement can be decomposed into a component along the critical direction and a component in the stable complementary subspace,
\begin{equation}
\mathbf{Y}-\mathbf{Y}_{c,i}
=
\boldsymbol{\psi}_i\mathcal{Y}_i
+
\boldsymbol{\eta}_i,
\qquad
\boldsymbol{\psi}_i^{\mathsf T}\boldsymbol{\eta}_i=0,
\label{eq:critical_stable_decomposition}
\end{equation}
where $\mathcal{Y}_i$ is the scalar amplitude along the critical eigenvector and $\boldsymbol{\eta}_i$ denotes the stable component.
Locally, the stable component is slaved to the critical amplitude and the loading parameter.
Thus, to leading order, the dynamics can be reduced to the critical direction, consistent with a Lyapunov--Schmidt reduction of the corresponding static equilibrium equations.
Introducing the local variables
\begin{equation}
\mathbf{Y}
=
\mathbf{Y}_{c,i}
+
\boldsymbol{\psi}_i\mathcal{Y}_i,
\qquad
\mathcal{M}_i
=
\Delta-\Delta_{c,i},
\qquad
\mathcal{T}_i
=
T-T_{c,i},
\label{eq:coupled_local_variables}
\end{equation}
and assuming a constant loading rate, the local distance from the static SN point evolves as
\begin{equation}
\mathcal{M}_i
=
\dot{\Delta}\mathcal{T}_i.
\end{equation}
Projecting the dimensionless equations of motion onto the critical eigenvector $\boldsymbol{\psi}_i$ then gives the local dynamic SN normal form
\begin{equation}
\frac{\mathrm{d}^{2}\mathcal{Y}_i}
{\mathrm{d}\mathcal{T}_i^{2}}
+
\Upsilon
\frac{\mathrm{d}\mathcal{Y}_i}
{\mathrm{d}\mathcal{T}_i}
=
\mathcal{A}_i\dot{\Delta}\mathcal{T}_i
+
\mathcal{B}_i\mathcal{Y}_i^{2}
+
\mathcal{O}
\left(
\mathcal{M}_i^{2},
\mathcal{M}_i\mathcal{Y}_i,
\mathcal{Y}_i^{3}
\right),
\label{eq:coupled_dynamic_sn_standard}
\end{equation}
where
\begin{equation}
\mathcal{A}_i
=
-
\boldsymbol{\psi}_i^{\mathsf T}
\frac{\partial \mathbf{G}}{\partial \Delta},
\qquad
\mathcal{B}_i
=
-
\frac{1}{2}
\boldsymbol{\psi}_i^{\mathsf T}
\frac{\partial^{2}\mathbf{G}}
{\partial\mathbf{Y}^{2}}
\left[
\boldsymbol{\psi}_i,
\boldsymbol{\psi}_i
\right].
\label{eq:coupled_normal_form_coefficients}
\end{equation}
The critical point $\Delta_{c,i}$, critical direction $\boldsymbol{\psi}_i$, and coefficients $\mathcal{A}_i$ and $\mathcal{B}_i$ all depend on the coupling stiffness $K_{12}^{s}$.
Here, ${\partial^{2}\mathbf{G}}/{\partial\mathbf{Y}^{2}}$ is a third-order tensor, and
${\partial^{2}\mathbf{G}}/{\partial\mathbf{Y}^{2}}
\left[
\boldsymbol{\psi}_i,
\boldsymbol{\psi}_i
\right]$
denotes its double contraction with $\boldsymbol{\psi}_i$, yielding a vector that represents the second directional derivative of $\mathbf{G}$ along the critical direction $\boldsymbol{\psi}_i$.
The two coupled SN branches are continued with respect to $K_{12}^{s}$ using a secant predictor--Newton corrector procedure applied to Eq.~\eqref{eq:coupled_sn_conditions}.
This branch-consistent continuation avoids unintended switching between nearby SN solutions and allows $\Delta_{c,i}$, $\boldsymbol{\psi}_i$, $\mathcal{A}_i$, and $\mathcal{B}_i$ to be tracked continuously as the coupling stiffness varies.
The local coefficients are evaluated analytically from Eq.~\eqref{eq:coupled_normal_form_coefficients}.

\subsubsection{Decoupled specialization}

The uncoupled limit of the general formulation is recovered by setting $K_{12}^{s}=0$.
In this limit,
\begin{equation}
\mathcal{U}_4(Y_1,Y_2;\Delta,0)
=
\widetilde{\mathcal{U}}_{4,1}(Y_1;\Delta)
+
\widetilde{\mathcal{U}}_{4,2}(Y_2;\Delta),
\label{eq:decoupled_energy_specialization}
\end{equation}
and the Hessian becomes diagonal.
The critical eigenvectors associated with the two independent units reduce to
\begin{equation}
\boldsymbol{\psi}_1
=
\begin{bmatrix}
1\\
0
\end{bmatrix},
\qquad
\boldsymbol{\psi}_2
=
\begin{bmatrix}
0\\
1
\end{bmatrix}.
\end{equation}
For either unit, introducing the generic coordinate $Y$, the reduced potential energy takes the form
\begin{equation}
\widetilde{\mathcal{U}}_4(Y;\Delta)
=
C_0+C_1Y+C_2Y^2+C_3Y^3+C_4Y^4,
\label{eq:decoupled_single_energy}
\end{equation}
with the corresponding unit-dependent parameters
\begin{equation}
(\kappa_s,\kappa_t,\alpha,Y)
=
\begin{cases}
(1,K_1^{t},\alpha_1,Y_1),
& \text{unit 1},\\[2mm]
(K_2^{s},K_2^{t},\alpha_2,Y_2),
& \text{unit 2}.
\end{cases}
\label{eq:decoupled_branch_parameters}
\end{equation}
The coefficients $C_i(\Delta)$ follow from Eqs.~\eqref{eq:C00}--\eqref{eq:C04} and retaining the terms associated with the selected unit.
Defining the scalar force residual
\begin{equation}
P(Y,\Delta)
=
\frac{\partial\widetilde{\mathcal{U}}_4}{\partial Y},
\label{eq:decoupled_force_residual}
\end{equation}
the SN point of the decoupled unit satisfies
\begin{equation}
P(Y_c,\Delta_c)=0,
\qquad
\frac{\partial P}{\partial Y}(Y_c,\Delta_c)=0.
\label{eq:decoupled_sn_conditions}
\end{equation}
Equation~\eqref{eq:coupled_normal_form_coefficients} then reduces exactly to
\begin{equation}
\mathcal{A}
=
-
\left.
\frac{\partial P}{\partial\Delta}
\right|_{(Y_c,\Delta_c)},
\qquad
\mathcal{B}
=
-
\frac{1}{2}
\left.
\frac{\partial^{2}P}{\partial Y^{2}}
\right|_{(Y_c,\Delta_c)}.
\label{eq:decoupled_AB_specialization}
\end{equation}

\subsubsection{Underdamped dynamics}
\label{sec:underdamped_dynamics}

\begin{figure}[!b]
    \centering
    \includegraphics[width=\linewidth]{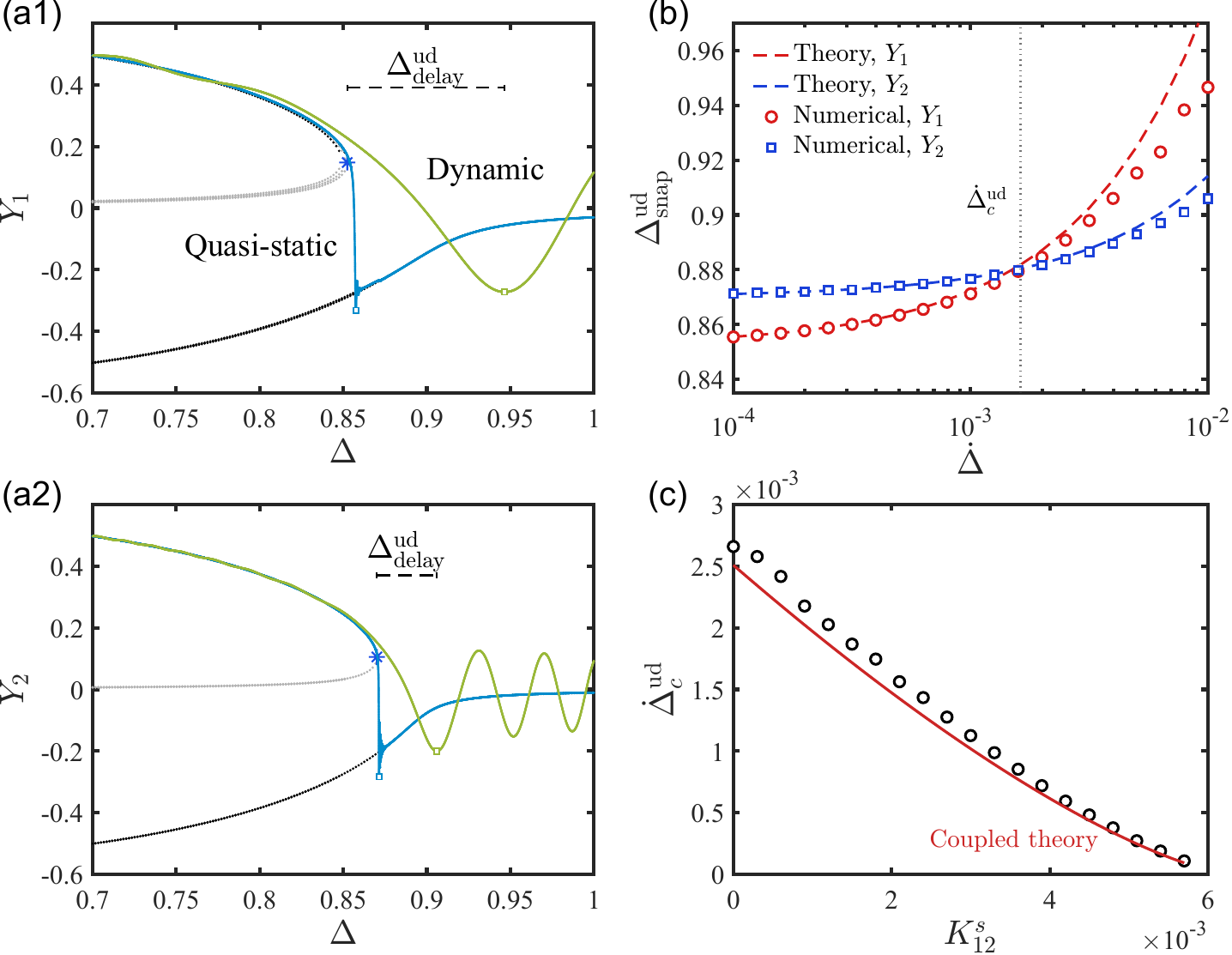}
    \caption{Underdamped dynamic snap-through and delayed instability.
    (a1,a2) Dynamic trajectories of units 1 and 2, showing delayed snap-through relative to the quasi-static instability point.
    Black and grey curves denote stable and unstable quasi-static equilibrium branches, respectively, while coloured curves denote representative complete dynamic trajectories.
    (b) Dynamically identified snap-through thresholds $\Delta_{\mathrm{snap}}^{\mathrm{ud}}$ as functions of $\dot{\Delta}$ at $K_{12}^{s}=0.002$.
    Dashed red and blue curves denote the coupled-theory predictions for units 1 and 2, respectively, while red circles and blue squares denote the corresponding full two-degree-of-freedom numerical results.
    The vertical grey dotted line marks the critical rate $\dot{\Delta}_{c}^{\mathrm{ud}}$.
    (c) Critical rate $\dot{\Delta}_{c}^{\mathrm{ud}}$ as a function of $K_{12}^{s}$.
    Open circles denote the full dynamic simulations, and the red solid curve denotes the coupled-theory prediction.}
    \label{fig:underdamping_plot}
\end{figure}

We next consider the underdamped regime, in which inertia dominates the local dynamics near the SN bottleneck.
For the $i$-th coupled SN point, we introduce the rescaled variables
\begin{equation}
\mathscr{Y}_i
=
\frac{\mathcal{Y}_i}
{\mathcal{Y}_{0,i}^{\mathrm{ud}}},
\qquad
\mathscr{T}_i
=
\frac{\mathcal{T}_i}
{\mathcal{T}_{0,i}^{\mathrm{ud}}},
\label{eq:coupled_underdamped_variables}
\end{equation}
where
\begin{equation}
\mathcal{T}_{0,i}^{\mathrm{ud}}
=
\left[
\frac{1}
{\left|\mathcal{A}_i\mathcal{B}_i\right|
\dot{\Delta}}
\right]^{1/5},
\qquad
\mathcal{Y}_{0,i}^{\mathrm{ud}}
=
\left|\mathcal{A}_i\right|
\dot{\Delta}
\left(
\mathcal{T}_{0,i}^{\mathrm{ud}}
\right)^3.
\label{eq:coupled_underdamped_scales}
\end{equation}
For the SN branches considered here,
$\mathcal{A}_i\mathcal{B}_i>0$.
Because the sign of $\boldsymbol{\psi}_i$ is arbitrary, it can be chosen such that $\mathcal{A}_i>0$ and $\mathcal{B}_i>0$.
{Under the above rescaling, the effective damping coefficient is $\Upsilon\mathcal{T}_{0,i}^{\mathrm{ud}}$.
In the underdamped limit, $\Upsilon\mathcal{T}_{0,i}^{\mathrm{ud}}\ll1$, the damping term is asymptotically negligible}, and Eq.~\eqref{eq:coupled_dynamic_sn_standard} reduce to
\begin{equation}
\frac{\mathrm{d}^{2}\mathscr{Y}_i}
{\mathrm{d}\mathscr{T}_i^{2}}
=
\mathscr{T}_i+\mathscr{Y}_i^{2}.
\label{eq:coupled_canonical_underdamped}
\end{equation}
The dynamically relevant solution is selected by matching to the stable quasi-static equilibrium branch,
\begin{equation}
\mathscr{Y}_i
\sim
-\sqrt{-\mathscr{T}_i},
\qquad
\mathscr{T}_i\rightarrow-\infty.
\label{eq:coupled_underdamped_matching}
\end{equation}
As the static SN point is approached, the stiffness associated with the critical mode vanishes and the intrinsic response timescale diverges.
The trajectory therefore does not depart immediately when $\Delta_{c,i}$ is crossed, but remains temporarily within the SN bottleneck.
The matched solution of Eq.~\eqref{eq:coupled_canonical_underdamped} escapes from this bottleneck at
\begin{equation}
\mathscr{T}_{\mathrm{esc}}^{\mathrm{ud}}
=
3.4039.
\label{eq:coupled_underdamped_escape}
\end{equation}
{giving the corresponding dimensionless local time delay}
\begin{equation}
\mathcal{T}_{\mathrm{delay},i}^{\mathrm{ud}}
=
3.4039\,
\mathcal{T}_{0,i}^{\mathrm{ud}}.
\end{equation}
Since $\Delta-\Delta_{c,i}=\dot{\Delta}\mathcal{T}_i$, the associated delay in the control parameter is
\begin{equation}
\Delta_{\mathrm{delay},i}^{\mathrm{ud}}
=
\Delta_{\mathrm{snap},i}^{\mathrm{ud}}
-
\Delta_{c,i}
=
3.4039
\left[
\frac{1}
{\left|\mathcal{A}_i\mathcal{B}_i\right|}
\right]^{1/5}
\dot{\Delta}^{4/5}.
\label{eq:coupled_underdamped_delay}
\end{equation}
Hence, the dynamically observed snap-through threshold is
\begin{equation}
\Delta_{\mathrm{snap},i}^{\mathrm{ud}}
=
\Delta_{c,i}(K_{12}^{s})
+
3.4039
\left[
\frac{1}
{
\left|
\mathcal{A}_i(K_{12}^{s})
\mathcal{B}_i(K_{12}^{s})
\right|
}
\right]^{1/5}
\dot{\Delta}^{4/5}.
\label{eq:coupled_underdamped_snap}
\end{equation}
Defining the coupling-dependent delay prefactor
\begin{equation}
C_i^{\mathrm{ud}}(K_{12}^{s})
=
3.4039
\left[
\frac{1}
{
\left|
\mathcal{A}_i(K_{12}^{s})
\mathcal{B}_i(K_{12}^{s})
\right|
}
\right]^{1/5},
\label{eq:coupled_underdamped_prefactor}
\end{equation}
the dynamic snap-through threshold can be written compactly as
\begin{equation}
\Delta_{\mathrm{snap},i}^{\mathrm{ud}}
=
\Delta_{c,i}
+
C_i^{\mathrm{ud}}\dot{\Delta}^{4/5}.
\label{eq:coupled_underdamped_competition}
\end{equation}
{The boundary between the two possible snapping orders occurs when the dynamically delayed thresholds coincide,}
\begin{equation}
\Delta_{\mathrm{snap},1}^{\mathrm{ud}}
=
\Delta_{\mathrm{snap},2}^{\mathrm{ud}},
\end{equation}
which gives the critical loading rate
\begin{equation}
\dot{\Delta}_{c}^{\mathrm{ud}}(K_{12}^{s})
=
\left[
\frac{
\Delta_{c,2}(K_{12}^{s})
-
\Delta_{c,1}(K_{12}^{s})
}{
C_1^{\mathrm{ud}}(K_{12}^{s})
-
C_2^{\mathrm{ud}}(K_{12}^{s})
}
\right]^{5/4}.
\label{eq:coupled_underdamped_critical}
\end{equation}

Figures~\ref{fig:underdamping_plot}(a1) and
\ref{fig:underdamping_plot}(a2) show that each trajectory remains near the SN bottleneck after the corresponding static threshold has been crossed, before eventually departing from the ghost of the vanished equilibrium.
This delayed departure is the dynamic signature of critical slowing down near an SN bifurcation.
Once the trajectory escapes from the bottleneck, the subsequent snap-through is rapid, with inertia dominating the local dynamics.
For the parameters considered here, $\Delta_{c,1}<\Delta_{c,2}$, so unit 1 snaps first in the quasi-static limit.
However, its underdamped delay prefactor is larger, $C_1^{\mathrm{ud}}>C_2^{\mathrm{ud}}$.
The static threshold separation and the rate-dependent delay terms in Eq.~\eqref{eq:coupled_underdamped_competition} therefore
favour opposite snapping orders.
At low loading rates, the static threshold separation dominates, giving $\Delta_{\mathrm{snap},1}^{\mathrm{ud}}<\Delta_{\mathrm{snap},2}^{\mathrm{ud}}$.
As the loading rate increases, the larger dynamic delay of unit 1 progressively compensates for its lower static threshold and eventually reverses the ordering, such that $\Delta_{\mathrm{snap},2}^{\mathrm{ud}}<\Delta_{\mathrm{snap},1}^{\mathrm{ud}}$.
The crossover in Fig.~\ref{fig:underdamping_plot}(b) therefore results from competition between two dynamically delayed local instabilities.
The coupled-theory predictions agree closely with the full two-degree-of-freedom simulations at low release rates, while the discrepancy increases as the rate increases.
This loss of accuracy is consistent with the local asymptotic nature of the SN reduction, since at higher rates the trajectory explores regions increasingly far from the immediate neighbourhood of the SN bottleneck.
Figure~\ref{fig:underdamping_plot}(c) shows that
$\dot{\Delta}_{c}^{\mathrm{ud}}$ decreases with increasing $K_{12}^{s}$ and approaches zero near the upper limit of the weak-coupling regime.
Elastic coupling modifies both the static threshold separation $\Delta_{c,2}-\Delta_{c,1}$ and the local normal-form coefficients $\mathcal{A}_i$ and $\mathcal{B}_i$, and hence the corresponding delay prefactors.
The coupled theory therefore captures the monotonic decrease and curvature of the numerically determined switching boundary.

\subsubsection{Overdamped dynamics}
\label{sec:overdamped_dynamics}

We next consider the overdamped regime, in which viscous resistance dominates the local escape dynamics near the coupled SN point.
For the $i$-th coupled SN point, we introduce the rescaled variables
\begin{equation}
\mathscr{Y}_i
=
\frac{\mathcal{Y}_i}
{\mathcal{Y}_{0,i}^{\mathrm{od}}},
\qquad
\mathscr{T}_i
=
\frac{\mathcal{T}_i}
{\mathcal{T}_{0,i}^{\mathrm{od}}},
\label{eq:coupled_overdamped_variables}
\end{equation}
where
\begin{equation}
\mathcal{T}_{0,i}^{\mathrm{od}}
=
\left[
\frac{\Upsilon^{2}}
{\mathcal{A}_i\mathcal{B}_i\dot{\Delta}}
\right]^{1/3},
\qquad
\mathcal{Y}_{0,i}^{\mathrm{od}}
=
\frac{
\mathcal{A}_i\dot{\Delta}
\left(
\mathcal{T}_{0,i}^{\mathrm{od}}
\right)^2
}
{\Upsilon}.
\label{eq:coupled_overdamped_scales}
\end{equation}
For the SN branches considered here, the orientation of
$\boldsymbol{\psi}_i$ is chosen, as in the preceding section, such that $\mathcal{A}_i>0$ and $\mathcal{B}_i>0$.
{Under the above rescaling, the relative magnitude of the inertial term is controlled by
$(\Upsilon\mathcal{T}_{0,i}^{\mathrm{od}})^{-1}$.
In the overdamped limit,
$\Upsilon\mathcal{T}_{0,i}^{\mathrm{od}}\gg1$, the inertial term is asymptotically negligible, and Eq.~\eqref{eq:coupled_dynamic_sn_standard} reduces to}
\begin{equation}
\Upsilon
\frac{\mathrm{d}\mathcal{Y}_i}
{\mathrm{d}\mathcal{T}_i}
=
\mathcal{A}_i\dot{\Delta}\mathcal{T}_i
+
\mathcal{B}_i\mathcal{Y}_i^{2}.
\label{eq:coupled_overdamped_sn}
\end{equation}
Substituting Eqs.~\eqref{eq:coupled_overdamped_variables} and \eqref{eq:coupled_overdamped_sn} gives
\begin{equation}
\frac{\mathrm{d}\mathscr{Y}_i}
{\mathrm{d}\mathscr{T}_i}
=
\mathscr{T}_i+\mathscr{Y}_i^{2}.
\label{eq:coupled_canonical_overdamped}
\end{equation}
The dynamically relevant solution is selected by matching to the stable quasi-static equilibrium branch,
\begin{equation}
\mathscr{Y}_i
\sim
-\sqrt{-\mathscr{T}_i},
\qquad
\mathscr{T}_i\rightarrow-\infty.
\label{eq:coupled_overdamped_matching}
\end{equation}
Using the Riccati--Airy transformation, the matched solution can be written as
\begin{equation}
\mathscr{Y}_i
=
\frac{\operatorname{Ai}'(-\mathscr{T}_i)}
{\operatorname{Ai}(-\mathscr{T}_i)}.
\label{eq:coupled_overdamped_airy_solution}
\end{equation}
where $\operatorname{Ai}$ denotes the Airy function of the first kind. 
The reduced solution becomes singular when the denominator first vanishes, corresponding to
\begin{equation}
\mathscr{T}_{\mathrm{esc}}^{\mathrm{od}}
=
2.3381.
\label{eq:coupled_overdamped_airy_zero}
\end{equation}

\begin{figure}[!b]
    \centering
    \includegraphics[width=\linewidth]{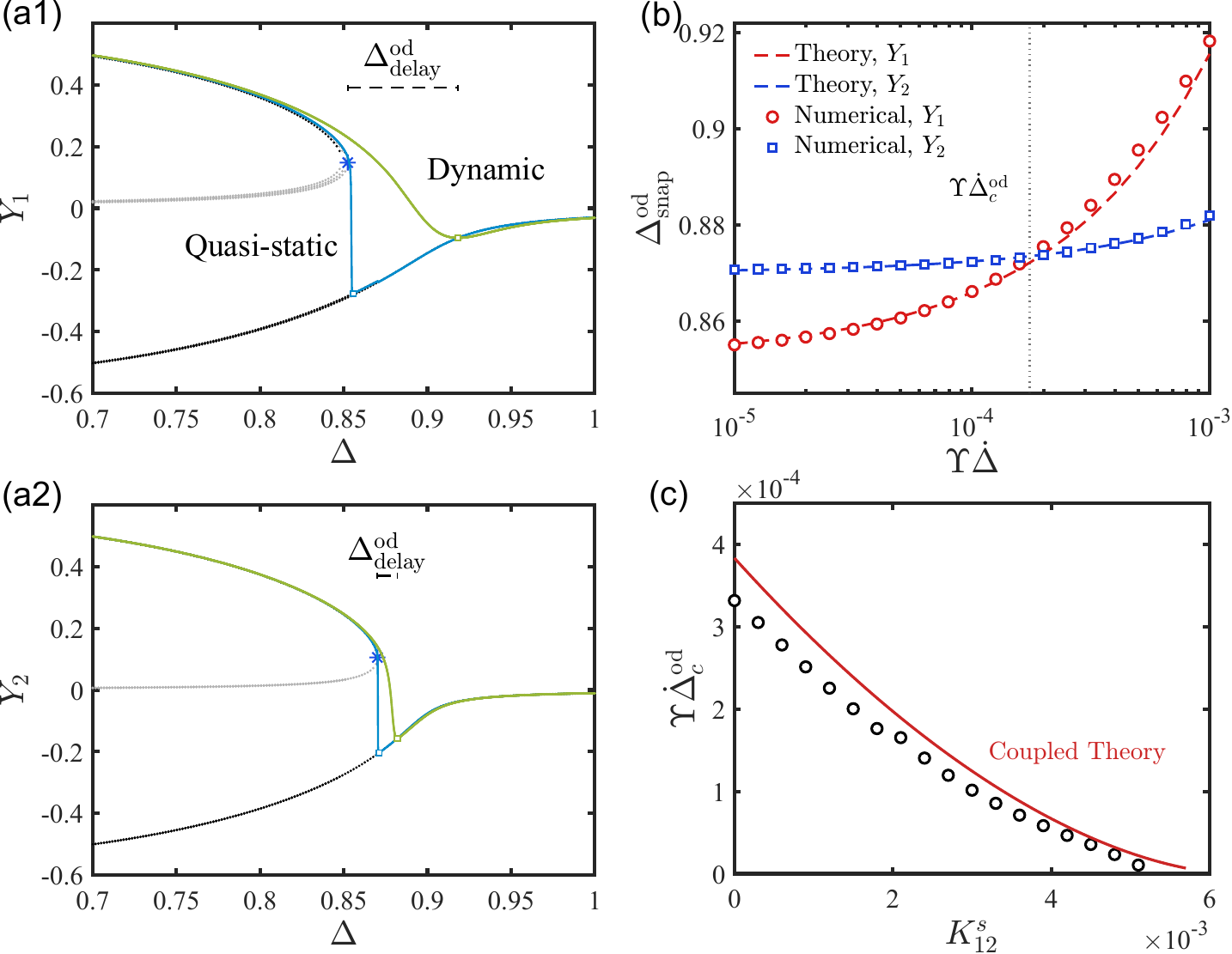}
    \caption{Overdamped dynamic snap-through and delayed instability.
    (a1,a2) Dynamic trajectories of units 1 and 2, showing delayed snap-through relative to the quasi-static instability point.
    Black and grey curves denote stable and unstable quasi-static equilibrium branches, respectively, while coloured curves denote representative complete dynamic trajectories.
    (b) Dynamically identified snap-through thresholds $\Delta_{\mathrm{snap}}^{\mathrm{od}}$ as functions of the scaled rate $\Upsilon\dot{\Delta}$ at $K_{12}^{s}=0.002$.
    Dashed red and blue curves denote the coupled-theory predictions for units 1 and 2, respectively, while red circles and blue squares denote the corresponding full two-degree-of-freedom numerical results.
    The vertical grey dotted line marks the critical scaled rate $\Upsilon\dot{\Delta}_{c}^{\mathrm{od}}$.
    (c) Critical scaled rate $\Upsilon\dot{\Delta}_{c}^{\mathrm{od}}$ as a function of $K_{12}^{s}$.
    Open circles denote the full dynamic simulations, and the red solid curve denotes the coupled-theory prediction.}
    \label{fig:overdamping_plot}
\end{figure}

This value characterises the asymptotic escape from the SN bottleneck when snap-through is identified by the singular departure of the local solution.
In the full numerical simulations, however, we operationally identify the snap-through point by the minimum displacement attained during the subsequent rapid transition.
Applying the same criterion to the universal reduced dynamics gives a corresponding prefactor of $3.31$, while leaving the asymptotic rate exponent unchanged.
The resulting delay in the control parameter is
\begin{equation}
\Delta_{\mathrm{delay},i}^{\mathrm{od}}
=
\Delta_{\mathrm{snap},i}^{\mathrm{od}}
-
\Delta_{c,i}
=
3.31
\left[
\frac{1}
{\left|\mathcal{A}_i\mathcal{B}_i\right|}
\right]^{1/3}
(\Upsilon\dot{\Delta})^{2/3},
\label{eq:coupled_overdamped_delay}
\end{equation}
and hence the dynamically observed snap-through threshold is
\begin{equation}
\Delta_{\mathrm{snap},i}^{\mathrm{od}}
=
\Delta_{c,i}(K_{12}^{s})
+
3.31
\left[
\frac{1}
{\left|\mathcal{A}_i(K_{12}^{s})
\mathcal{B}_i(K_{12}^{s})\right|}
\right]^{1/3}
(\Upsilon\dot{\Delta})^{2/3}.
\label{eq:coupled_overdamped_snap}
\end{equation}
Defining the coupling-dependent overdamped delay prefactor
\begin{equation}
C_i^{\mathrm{od}}(K_{12}^{s})
=
3.31
\left|
\mathcal{A}_i(K_{12}^{s})
\mathcal{B}_i(K_{12}^{s})
\right|^{-1/3}.
\label{eq:coupled_overdamped_prefactor}
\end{equation}
The boundary between the two possible snapping orders occurs when
$\Delta_{\mathrm{snap},1}^{\mathrm{od}}
=
\Delta_{\mathrm{snap},2}^{\mathrm{od}}$,
which gives the critical scaled loading rate
\begin{equation}
\Upsilon\dot{\Delta}_{c}^{\mathrm{od}}(K_{12}^{s})
=
\left[
\frac{
\Delta_{c,2}(K_{12}^{s})
-
\Delta_{c,1}(K_{12}^{s})
}{
C_1^{\mathrm{od}}(K_{12}^{s})
-
C_2^{\mathrm{od}}(K_{12}^{s})
}
\right]^{3/2}.
\label{eq:coupled_overdamped_critical}
\end{equation}

Figures~\ref{fig:overdamping_plot}(a1) and
\ref{fig:overdamping_plot}(a2) show that each trajectory remains near the SN bottleneck after the corresponding static threshold has been crossed, before eventually departing from the ghost of the vanished equilibrium.
This delayed departure is the dynamic signature of critical slowing down near an SN bifurcation.
In contrast to the underdamped regime, escape from the local bottleneck is governed by viscous drift rather than inertial acceleration.
For the parameters considered here, $\Delta_{c,1}<\Delta_{c,2}$, so unit 1 snaps first in the quasi-static limit.
However, its overdamped delay prefactor is larger, $C_1^{\mathrm{od}}>C_2^{\mathrm{od}}$.
The static threshold separation and the rate-dependent delay terms in Eq.~\eqref{eq:coupled_overdamped_snap} therefore favour opposite snapping orders.
At low scaled release rates, the static threshold separation dominates, giving $\Delta_{\mathrm{snap},1}^{\mathrm{od}}<\Delta_{\mathrm{snap},2}^{\mathrm{od}}$.
As the scaled release rate increases, the larger dynamic delay of unit 1 progressively compensates for its lower static threshold and eventually reverses the ordering, such that $\Delta_{\mathrm{snap},2}^{\mathrm{od}}<\Delta_{\mathrm{snap},1}^{\mathrm{od}}$.
The crossover in Fig.~\ref{fig:overdamping_plot}(b) therefore results from competition between two dynamically delayed local instabilities.
Over the scaled-rate range considered, the coupled-theory predictions remain in close agreement with the full two-degree-of-freedom simulations and accurately capture the crossover of the two dynamic snap-through thresholds.
Figure~\ref{fig:overdamping_plot}(c) shows that
$\Upsilon\dot{\Delta}_{c}^{\mathrm{od}}$ decreases with increasing $K_{12}^{s}$ and approaches zero as the two quasi-static instability thresholds approach one another.
Elastic coupling modifies both the static threshold separation
$\Delta_{c,2}-\Delta_{c,1}$ and the local normal-form coefficients $\mathcal{A}_i$ and $\mathcal{B}_i$, and hence the corresponding overdamped delay prefactors.
The resulting coupled theory captures the monotonic decrease and curvature of the numerically determined switching boundary.

\subsection{Comparison of the two asymptotic regimes}
\label{sec:Comparison}

The two asymptotic limits differ in their local escape dynamics:
inertia gives $\Delta_{\mathrm{delay}}^{\mathrm{ud}}\sim\dot{\Delta}^{4/5}$,
whereas viscous drift gives
$\Delta_{\mathrm{delay}}^{\mathrm{od}}\sim(\Upsilon\dot{\Delta})^{2/3}$.
Despite these different scaling laws, the underlying pathway-selection mechanism is the same.
Dynamic bifurcation delay shifts the observed snap-through thresholds away from their quasi-static values, and unequal delays can reverse the ordering of competing local instabilities.
Consequently, the unit that loses stability first in the quasi-static limit may snap later under finite-rate loading.
Both underdamped and overdamped bifurcation delays can therefore reorder competing snap-through events and switch the selected transition pathway.

\section{Discussion}
\label{sec:Discussion}

The present two-mass von Mises truss provides a reduced-order representation of a broader class of beam-, arch-, and shell-based multistable systems, including those illustrated in Fig.~\ref{fig:bistable_plot}.
A snapping beam, arch, or shell may be represented locally by a generalised coordinate associated with its dominant saddle-node (SN) instability, while elastic interactions transmitted through shared supports, compliant frames, interconnecting ligaments, or elastic substrates provide physical counterparts of the coupling stiffness in the present model.
Such a reduction is consistent with previous studies showing that a Mises truss can reproduce the principal snap-through regimes of a continuous arch \citep{gomez2019dynamics}, that the dynamics of a continuous elastic arch near a slowly traversed SN bifurcation reduce to a generic low-dimensional normal form \citep{Liu2021Delayed}, and that a two-mass von Mises truss can capture rate-dependent transition pathways observed experimentally in coupled snapping arches \citep{Wang2024Transient}.
These correspondences provide a direct route towards experimental realisation using coupled bistable beams, arches, or shells: the geometry and local stiffness of each element set its individual instability threshold, the connecting structure controls the elastic coupling, and the imposed actuation rate provides the dynamic control parameter.

Beyond this structural correspondence, the present results provide a simple physical interpretation of pathway selection in multistable systems.
In the quasi-static limit, the selected pathway is governed by the relative ordering of competing local instability thresholds together with the accessibility of the remaining stable states.
Elastic coupling modifies these thresholds and can therefore reorganise the sequence of local instabilities, switching the system between sequential and cooperative pathways.
At finite loading rates, each local instability acquires its own dynamic bifurcation delay, so the dynamically observed snap-through thresholds need not preserve their quasi-static ordering.
A pathway can therefore be reordered without any change to the underlying equilibrium bifurcation structure.
This distinction reveals two complementary mechanisms for pathway control: structural coupling reorganises the quasi-static instability landscape, whereas loading rate dynamically reorders the effective transition thresholds.
From this viewpoint, rate-controlled pathway switching can be interpreted as a competition between multiple delayed local instabilities, providing a natural basis for extending the framework to larger assemblies of interacting multistable elements \citep{Shohat2025}.

For multistable metamaterials and thin-walled structures such as those represented in Fig.~\ref{fig:bistable_plot}(b), this mechanism suggests new opportunities for controlling not only which stable state is reached, but also how that state is reached.
Elastic coupling and loading rate could, for example, be used jointly to programme sequential or cooperative switching in mechanical metamaterials, to create adaptive structures with different responses under slow and rapid loading, or to control the order of local shape changes in deployable systems.
The present reduction is expected to be most applicable when the relevant transitions are governed by well-separated SN bifurcations and the dynamics are dominated by a small number of structural coordinates.
When several deformation modes interact strongly, when the system moves far from the local SN bottleneck, or when other bifurcation types become important, additional coordinates or different local normal forms may be required.
For example, recent work on shallow arches has shown that pitchfork-type bifurcation buckling can introduce precursor oscillations and strong sensitivity of the snap-through time to imperfections \citep{Simpkins2026}.
Extending the present framework to multimode systems, mixed bifurcation structures, and larger networks of coupled multistable elements therefore represents an important direction for future work.

\section{Conclusion}
\label{sec:conclusion}

We have investigated how elastic coupling and loading rate jointly govern transition-pathway selection in a minimal model of coupled multistable snap-through.
Under quasi-static release, varying the coupling stiffness reorganises the relative ordering of competing SN bifurcations, thereby selecting between two sequential pathways through distinct mixed states and a direct cooperative transition.
Under finite-rate release, the two local instabilities acquire unequal dynamic bifurcation delays, so their dynamically observed snap-through thresholds can cross even when their quasi-static ordering remains unchanged.
The unit that loses stability first in the quasi-static limit may therefore snap later dynamically, with the delay following the characteristic $4/5$ and $2/3$ scaling laws in the underdamped and overdamped regimes, respectively.

A coupling-dependent local theory was developed by continuously tracking the coupled SN branches and reducing the dynamics near each SN point along its critical mode.
The resulting normal-form theory predicts the dynamic snap-through thresholds and the critical rate for pathway switching, while explicitly accounting for the dependence of the local instability dynamics on elastic coupling.
The predictions agree closely with the full two-degree-of-freedom simulations in the overdamped regime and at low release rates in the underdamped regime.
The increasing discrepancy at higher underdamped rates reflects the local asymptotic nature of the reduction, as the trajectory increasingly explores regions beyond the immediate SN bottleneck.

Together, these results show that transition pathways constitute a controllable feature of multistable dynamics rather than a fixed consequence of the quasi-static bifurcation diagram.
Elastic coupling sets the relative ordering of competing quasi-static instabilities, while loading rate can dynamically reorder their effective snap-through thresholds.
Their interplay therefore provides a simple mechanism for selecting distinct transition pathways between the same initial and final states without altering the underlying structural architecture.

\section*{Acknowledgments}
{W. Huang acknowledges the start-up funding from Newcastle University, UK. J. Zhang and K. Huang acknowledge support from the National Natural Science Foundation of China (Grant Nos. 92271104 and 12102017). K. Huang also acknowledges support from the Academic Excellence Foundation of Beihang University for PhD Students. M. Liu acknowledges the start-up funding from The University of Birmingham, UK.}

\appendix
\renewcommand{\thesection}{Appendix \Alph{section}}

\section{Prefactor in overdamped dynamic snap-through}
\label{sec:AppendixA}

\begin{figure}[!h]
    \centering
    \includegraphics[width=\linewidth]{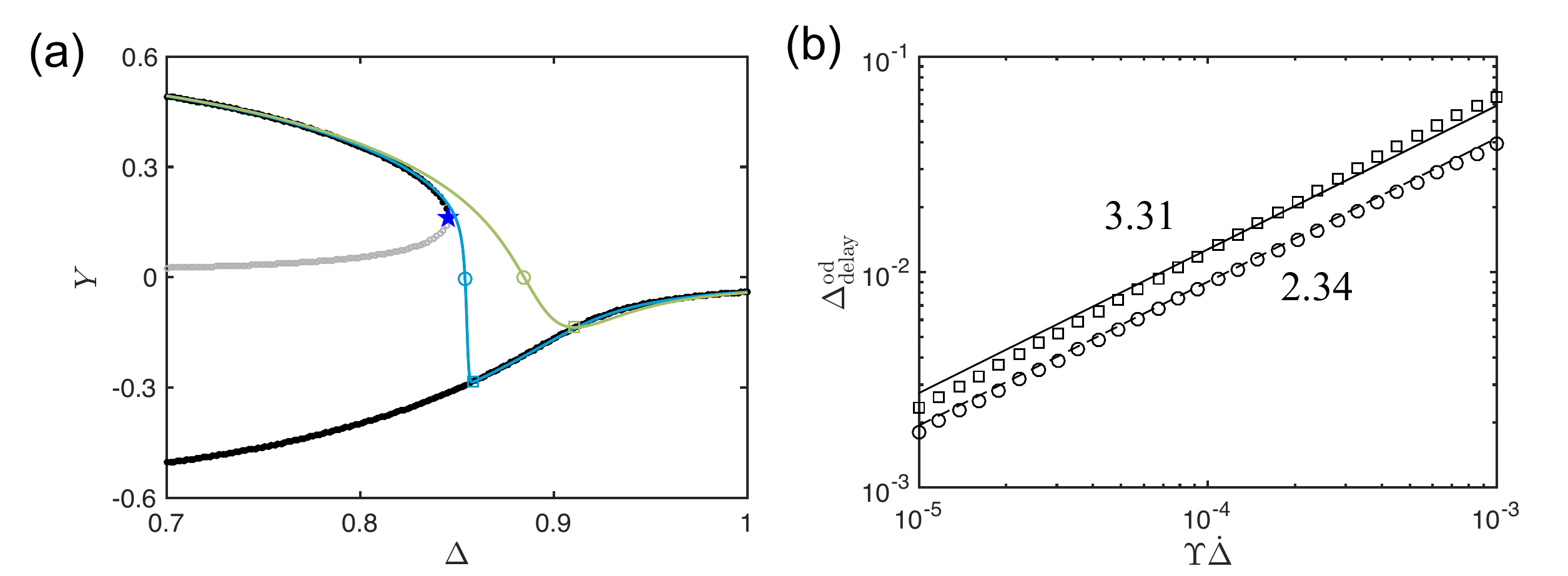}
    \caption{Prefactors for overdamped dynamic snap-through delay.
    (a) Numerical identification of the snap-through point using two criteria: the crossing of $Y=0$ (circle) and the minimum displacement along the dynamic trajectory (square).
    (b) Dynamic delay $\Delta_{\mathrm{delay}}^{\mathrm{od}}$ as a function of the scaled rate $\Upsilon\dot{\Delta}$ for the two criteria.
    Symbols denote numerical results, while solid lines show the corresponding theoretical scaling predictions, with prefactors $2.34$ and $3.31$, respectively.}
\label{fig:od_prefactor_plot}
\end{figure}

This appendix examines the prefactor used to define the delayed snap-through point in the overdamped regime.
The corresponding numerical validation is presented in Fig.~\ref{fig:od_prefactor_plot}.
For overdamped dynamics, the theoretical delay is determined by the escape point of the reduced normal-form equation.
The associated prefactor is given by the magnitude of the first zero of the Airy function, $\mathscr{T}_{0}^{\mathrm{od}} \simeq 2.34$.
This prefactor applies when the snap-through point is defined as the instant at which the trajectory crosses $Y=0$, indicated by the circular marker in Fig.~\ref{fig:od_prefactor_plot}(a).
In the present study, however, the snap-through point is defined by the minimum displacement attained along the trajectory, as indicated by the square marker in Fig.~\ref{fig:od_prefactor_plot}(a).
Under this definition, the numerical results are best 
described by the fitted prefactor $\mathscr{T}_{0}^{\mathrm{od}} = 3.31$.
Importantly, changing the snap-through criterion affects only the numerical prefactor, while the scaling law remains unchanged: $\Delta_{\mathrm{delay}}^{\mathrm{od}} \sim (\Upsilon \dot{\Delta})^{2/3}. $
Figure~\ref{fig:od_prefactor_plot}(b) compares the delayed displacement obtained from numerical simulations with the corresponding theoretical predictions using both prefactors.

\section{Video}
\label{sec:AppendixC}

We provide two videos as supplementary material to illustrate our dynamic results in Fig.~\ref{fig:dynamic_plot}.

\bibliographystyle{elsarticle-harv}
\bibliography{paper}

\end{document}